\documentclass[conf]{new-aiaa}
\usepackage[utf8]{inputenc}
\usepackage{multirow}
\usepackage{graphicx}
\usepackage{amsmath}
\usepackage[version=4]{mhchem}
\usepackage{siunitx}
\usepackage{longtable,tabularx}
\usepackage[dvipsnames]{xcolor}
\def\drwln#1#2{\raise 2.5pt\vbox{\hrule width #1pt height #2pt}}
\def\solid{\ \drwln{20}{1.0}\ }
\def\spc#1{\hskip #1pt}
\def\dashed{\ \hbox {\drwln{4}{1.0}\spc{2}
                   \drwln{4}{1.0}\spc{2}\drwln{4}{1.0}}\nobreak\ }

\def\dotted{\hbox {\drwln{1}{1.0}\spc{2}
                   \drwln{1}{1.0}\spc{2}\drwln{1}{1.0}\spc{2}\drwln{1}{1.0}\spc{2}\drwln{1}{1.0}}\nobreak\ }

\title{Separation of a Laminar Boundary Layer Subjected to Pressure Gradients with Spanwise Variations}

\author{John Marshall Cooper \footnote{Undergraduate Student}, Benjamin S. Savino\footnote{Ph.D. Candidate}, Benjamin Kellum Cooper \footnote{Undergraduate Student}, and Wen Wu\footnote{Assistant Professor and AIAA member. Corresponding author. Email address: wu@olemiss.edu}}
\affil{Department of Mechanical Engineering, University of Mississippi, University, MS, 38677, USA}

\begin{document}

\maketitle

\section{Introduction}
Non-uniform adverse pressure gradients (APGs) and three-dimensional laminar separation bubbles (LSBs) are ubiquitous in engineering settings. For example, wind turbine blades often feature variable chords, airfoil cross-sections, and twist angles along their span, resulting in variations of flow quantities in the spanwise direction \citep{Schubel12}. Similarly, gas turbines experience highly three-dimensional flow due to the complex geometry of the blades, including three-dimensional separation near the leading edge of the blade and the base wall \citep{Langston77, Hah84}. Further, the tip vortices of finite-aspect-ratio wings have been shown to lead to spanwise variations of flow quantities and LSBs \citep{Taira09, Devoria17, Zhu23}.

Despite the prevalence of three-dimensional flows in engineering, studies performed to characterize the behavior and influence of LSBs often resort to two-dimensional configurations. For experimental works, this implies that the aspect ratio of the test section is sufficient such that side-wall effects are minimal, mimicking an infinite span \cite{Horton68, Omeara87}. For computational studies, two-dimensionality is ensured through employing a spanwise homogeneous geometry or APG, along with the use of periodic boundary conditions in the spanwise direction \citep{Pauley90,Alam00}. While the cited studies, among a plethora of others, have provided valuable insight into the behavior of two-dimensional LSBs, three-dimensional ones are far less investigated.

The majority of literature on three-dimensional LSBs emphasizes the effect of tip vortices produced by finite-aspect-ratio wings. Significant progress has been made in explaining the three-dimensionalization of LSBs over various wing configurations. Results have shown alteration of vortex evolution and shedding, modulation of wake stability and structure, as well as the influences these changes have on wing performance (i.e., lift curves, drag curves, etc.) \citep{Bastedo86, Torres04, Taira09, Zhang20a, Zhang20b, Toppings21, Zhu23}. That said, the observed three-dimensional behaviors are a product of the interaction of various coherent structures, including the LSB itself, streamwise-oriented tip vortices, spanwise-oriented Kelvin-Helmholtz vortices, von K\'arm\'an vortex street, etc. Thus, the results are strongly dependent on flow configuration (airfoil shape, planform shape, aspect ratios). 

This study aims to isolate the effect of non-uniform APGs on LSBs, an area to the best of our knowledge which has not been extensively addressed.  Specifically, the primary objectives of the following manuscript are: 1) identify key characteristics of pressure-induced three-dimensional LSBs, and how these vary from their two-dimensional counterpart; 2) identify how three-dimensional characteristics change with APG non-uniformity and APG strength; and 3) determine how non-uniform APGs introduce three-dimensionality to LSBs.  To do this, flat-plate boundary layers, which eliminate the geometry dependence, with various APG configurations are compared. In the following, we discuss methodology, followed by mean flow fields and turbulence statistics. Instantaneous flow fields are provided for insights into physical mechanisms responsible for observed changes in statistics.

\section{Methodology}
The incompressible Navier-Stokes equations normalized by the inflow boundary layer thickness ($\delta$) and free stream velocity ($U_{\infty}$), given in Eq. (\ref{eqn:NS}), are solved via direct simulation (DNS). The Reynolds number is $Re = \delta U_{\infty}/\nu = 1000$. The computational domain is 90$\delta$ $\times$ 15$\delta$ $\times$ 24$\delta$ in the streamwise ($x$), wall-normal ($y$), and spanwise ($z$) directions. The domain is discretized by $N_i \times N_j \times N_k = 960 \times 199 \times 192$ points in the $x$, $y$, and $z$ directions.

 The computational solver is a well-validated, in-house program that utilizes a second-order accurate central difference scheme for all spatial derivatives, with variables defined on a staggered grid \citep{Keating04,Wu19, Wu24}. Semi-implicit time advancement is employed, with a second-order Adams-Bashforth method for convective terms and an implicit Crank-Nicolson method for the viscous terms. Time is advanced with a fractional-step algorithm. The Poisson equation is solved using a pseudo-spectral method, employing a cosine transform in the streamwise direction and a fast Fourier transform in the spanwise direction, followed by a direct solve of the resulting tri-diagonal matrix \citep{Moin2010}. 

\begin{equation}
    \frac{\partial u_j}{\partial x_j} = 0; \;\;
    \frac{\partial u_i}{\partial t} + \frac{\partial u_i u_j}{\partial x_j} = - \frac{\partial P}{\partial x_i} + \frac{1}{Re} \frac{\partial^2 u_i}{\partial x_j^2}
    \label{eqn:NS}
\end{equation}

\begin{figure}
    \centering
    \includegraphics[width=0.49\textwidth]{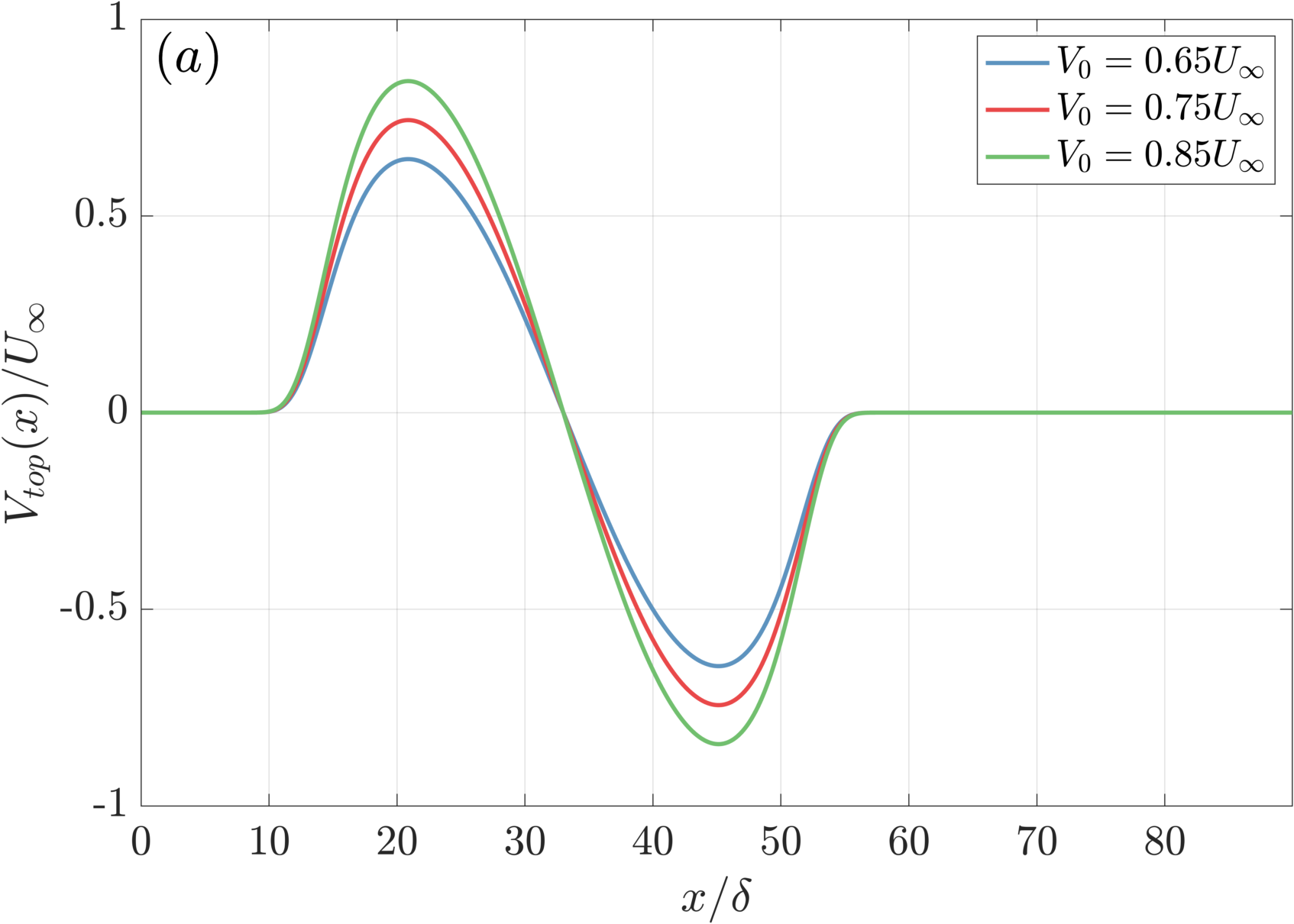}
    \includegraphics[width=0.49\textwidth]{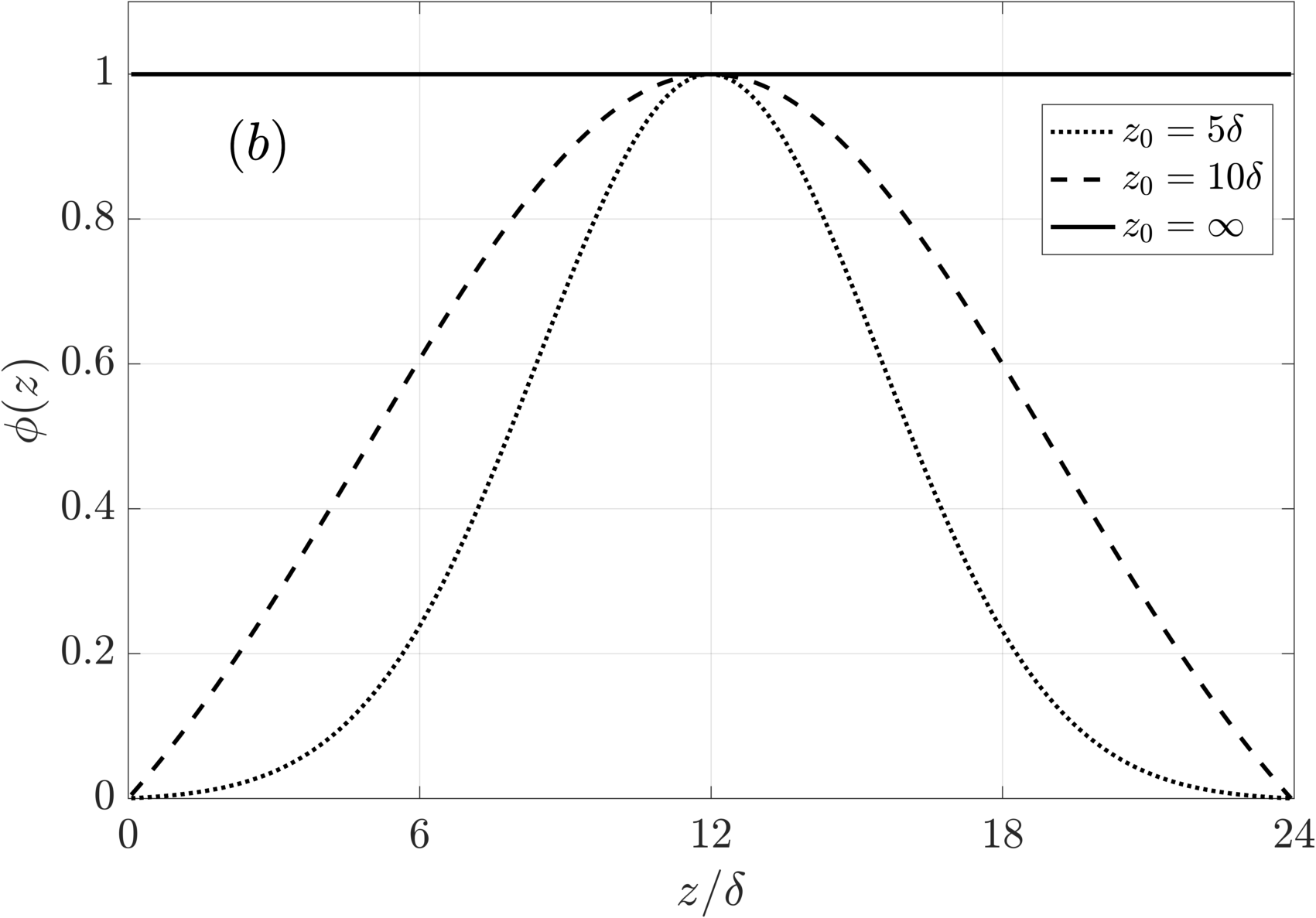}
    \caption{(a) Streamwise profile of $V_{top}$. (b) Weighting function utilized to introduce spanwise-dependence of suction boundary condition.}
    \label{fig:Vtop}
\end{figure}

\begin{figure}[h!]
    \centering
    \includegraphics[width=1.0\textwidth]{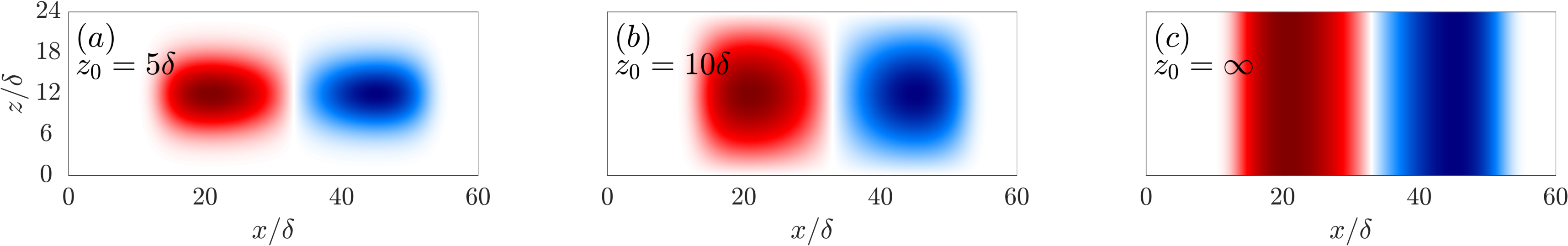}
    \includegraphics[width=0.2\textwidth]{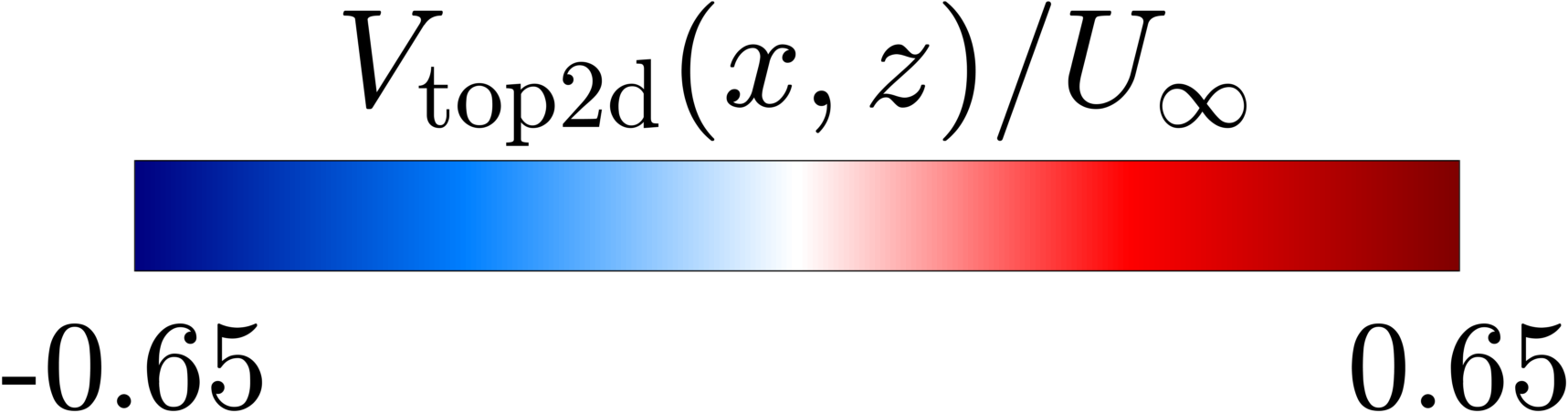}
    \caption{Applied wall-normal velocity profile ($V/U_{\infty}$) on the top boundary. Note only the suction level $V_0=0.65U_{\infty}$ is shown. Subplots (a), (b), and (c) display $V_\text{top2d}(x,z)$ for $z_0 = 5\delta$, $10\delta$, and $\infty$, respectively.}
    \label{fig:Vtop2d}
\end{figure}

To generate an APG and induce a finite-length LSB, we apply a suction-blowing profile to the top boundary. Three suction strengths are tested. The streamwise-dependent suction-blowing profiles are shown in Fig. \ref{fig:Vtop}(a). The one-dimensional $V_\text{top}(x)$ is scaled by a spanwise distribution to introduce spanwise variation to the APG. Three spanwise distributions are tested, as shown in Fig. \ref{fig:Vtop}(b).
The streamwise profile is multiplied with the spanwise distribution given to generate the suction-blowing distribution that is a function of $x$ and $z$:
\begin{equation}
    V_\text{top2d}(x,z) = \phi(z)V_\text{top}(x)
\end{equation}
The two-dimensional maps of the imposed top boundary condition ($V_\text{top2d}(x,z)$) are shown in Fig. \ref{fig:Vtop2d} for reference. To emphasize the spanwise variation dictated by $z_0$, only one level of suction is shown ($V_0 = 0.65U_{\infty}$).
The remaining boundary conditions are as follows: on the top boundary, streamwise velocity ($u$) is set to enforce zero spanwise vorticity, and a zero wall-normal gradient of the spanwise velocity ($w$) is imposed. A Blasius laminar boundary layer profile is set at the inflow, and a convective boundary condition is used at the outflow. The no-slip condition is enforced on the bottom wall. Finally, a periodic boundary condition is employed in the spanwise direction.

To differentiate between the nine cases resulting from the variation of $V_0$ and $z_0$, the following nomenclature is adopted: each case is denoted as `zXXXvYYY', where `XXX' is either `5', `10', or `inf', corresponding to $z_0 = 5\delta$, $10\delta$, and $\infty$, respectively. `YYY' is either `065', `075', or `085', corresponding to $V_0 = 0.65U_{\infty}$, $0.75U_{\infty}$, and $0.85U_{\infty}$. Throughout this abstract, the terms `non-uniform' and `three-dimensional' are used interchangeably to describe cases where $z_0 = 5\delta$ and $z_0 = 10\delta$ (i.e., when the APG is not applied uniformly across the domain span). `Two-dimensional' is used to describe cases where $z_0=\infty$ (APG applied uniformly across the span). Similarly, `suction strength,' `suction level,' and `APG strength' are used interchangeably to reference the level of $V_0/U_{\infty}$. After the statistically steady state is reached in each case, statistics are sampled every $9 \delta/U_{\infty}$ for a total of $1500 \delta/U_{\infty}$, ensuring statistically converged results. 
In the following discussion, capital letters ($U$) or $\overline{\left( . \right)}$ indicate ensemble (time) averaged quantifies. $\left< .\right>$ indicates quantities averaged over the spanwise ($z$) direction. $\left( .\right)'$ indicates fluctuations from the mean, and lowercase variables ($u_i$) indicate instantaneous quantities. 
\section{Results}
\subsection{Mean flow}
\begin{figure}[h!]
    \centering
    \includegraphics[width=0.35\textwidth]{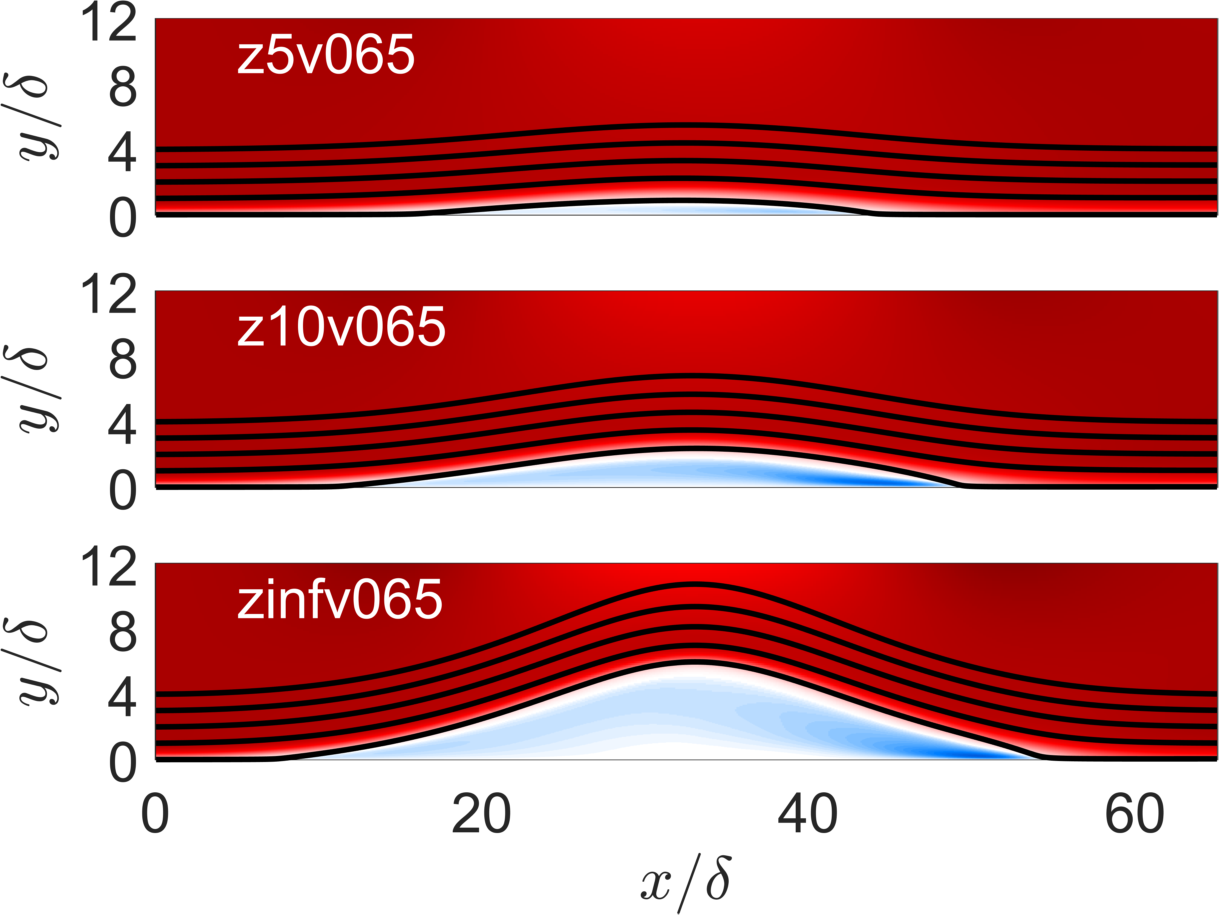}
    \includegraphics[width=0.31\textwidth]{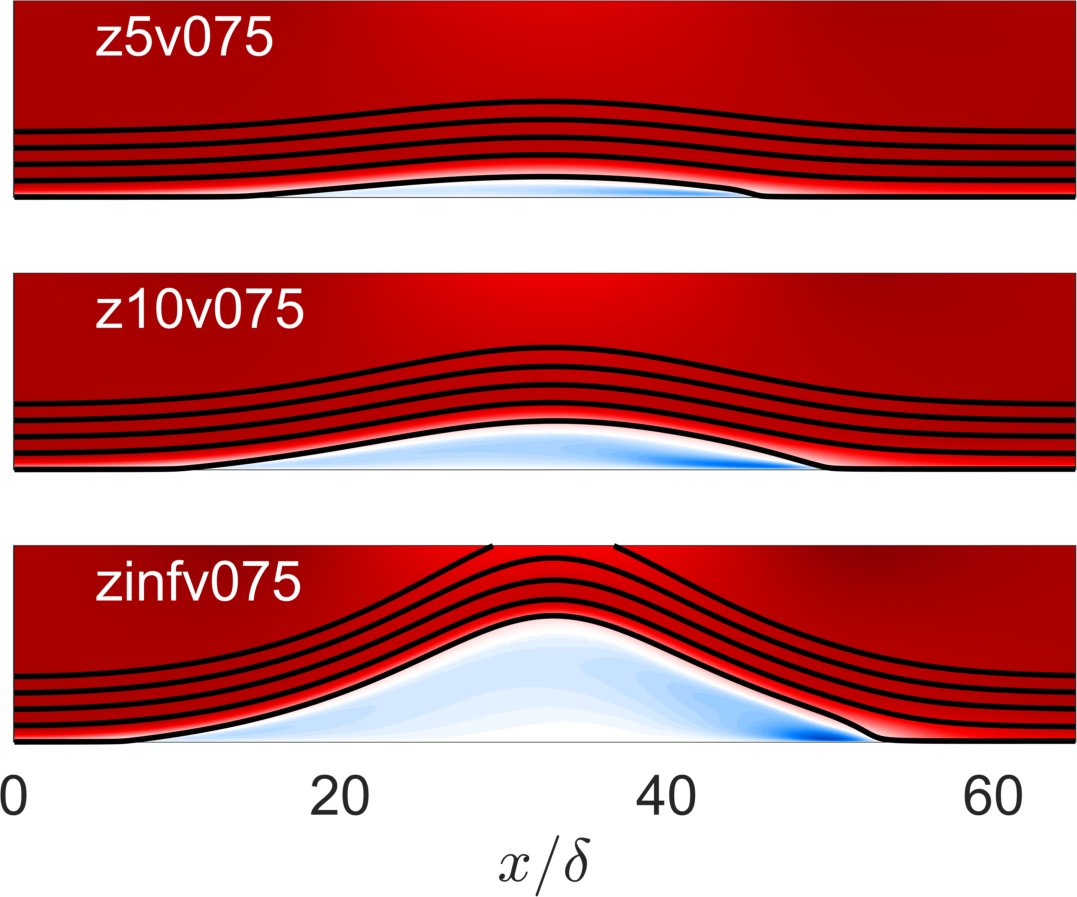}
     \includegraphics[width=0.31\textwidth]{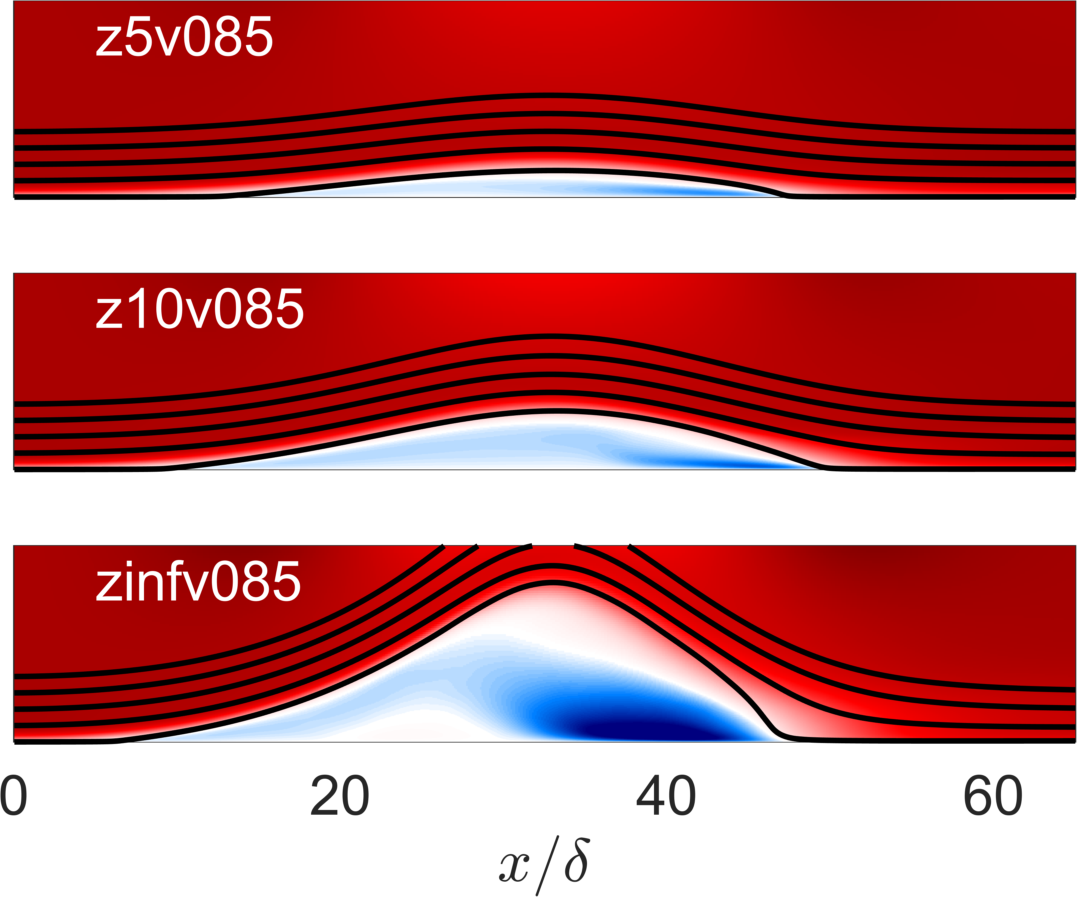}
     \hspace*{0.8cm}
     \includegraphics[width=0.15\textwidth]{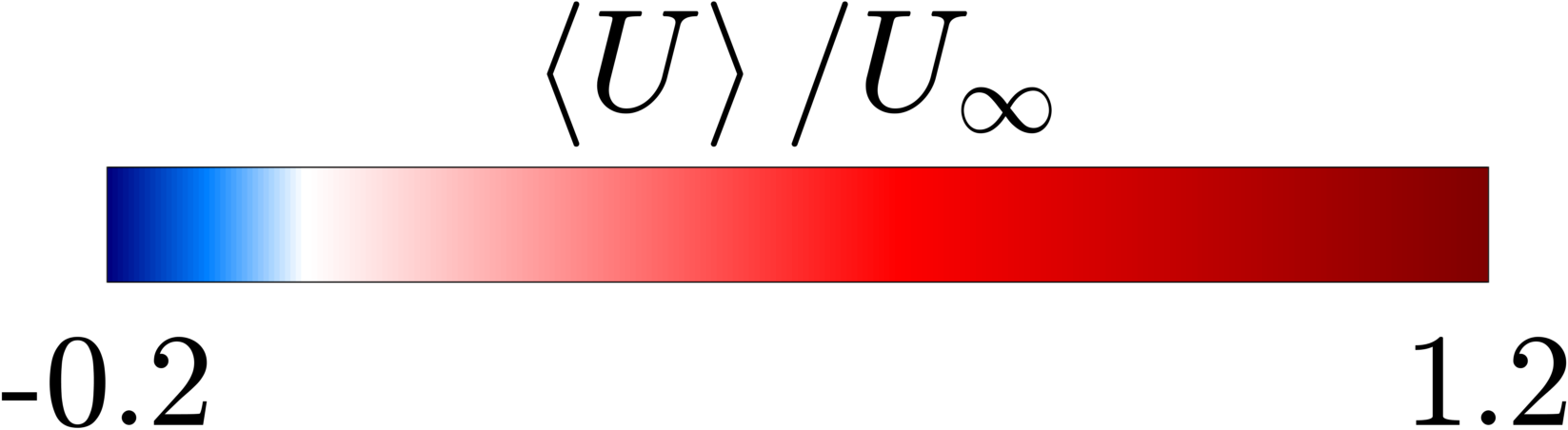}
     \caption{Time and spanwise averaged streamwise velocity fields with select mean streamlines starting from $x/\delta =0$ and $y/\delta = 0.004,\; 1.0,\; 2.0,\; 3.0,\; 4.0$. Rows correspond to suction width, while columns correspond to suction strength. }
     \label{fig:Uxy}
\end{figure}

Figure \ref{fig:Uxy} shows the mean streamwise velocity. Note that it is also averaged over the span, which allows for comparison with conventional two-dimensional LSBs.
The time and spanwise averaged separation bubble clearly shows a strong dependence on both suction strength (and the resultant APG) and the suction width in the span. As APG increases, the height of the separation bubble increases, and so does the magnitude of the reverse flow near reattachment. Similarly, as suction width increases, both height and reverse flow magnitude increase. Increasing suction width also results in steeper streamlines on the lee side of the LSB. This is most prominent in cases zinfv075 and zinfv085. 
Changes to the mean separation bubble are more drastic with changing suction width as opposed to changing suction strength.
 
\begin{figure}[h!]
    \centering
    \includegraphics[width=0.55\textwidth]{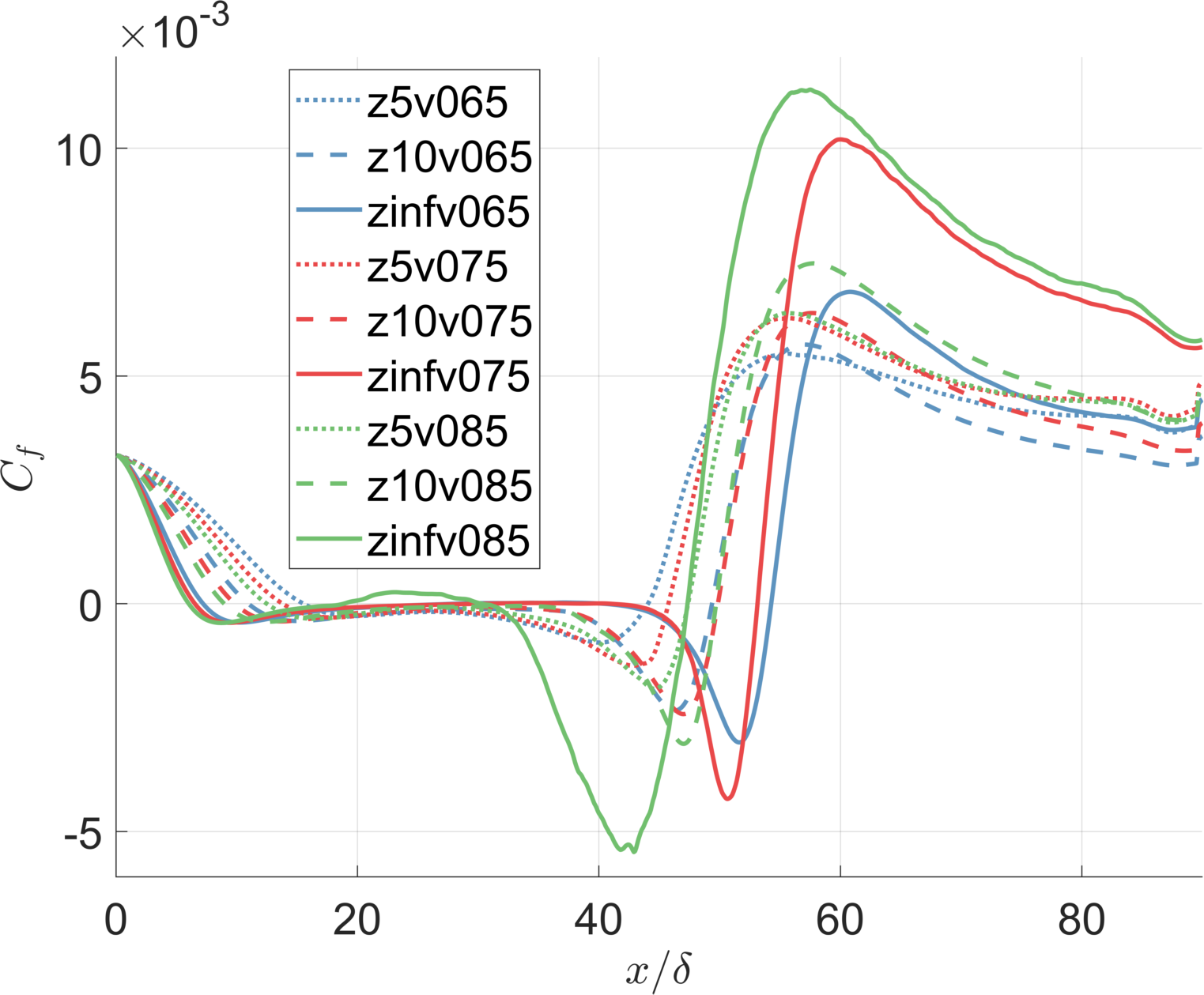}
    \caption{Time and spanwise averaged skin-friction coefficient. Line color corresponds to suction  strength: {\color{RoyalBlue}\solid}$V_0 = 0.65U_{\infty}$; {\color{BrickRed}\solid} $V_0 = 0.75U_{\infty}$; {\color{ForestGreen}\solid} $V_0 = 0.85U_{\infty}$. Line style corresponds to suction width: \dotted $z_0 = 5\delta$; \dashed $z_0 = 10\delta$; \solid $z_0 = \infty$.}
    \label{fig:Cf}
\end{figure}

The mean skin-friction coefficient ($C_f = 2\tau_w/ U_{\infty}^2$, $\tau_w = 1/Re \partial \left< U \right >/\partial y$) is plotted in Fig. \ref{fig:Cf}. Color corresponds to suction strength while line style corresponds to spanwise suction width. The behavior at the separation point (defined as the $x$ location where $C_f = 0 $ and $\partial C_f/ \partial x < 0$) is evident: stronger APG leads to earlier separation. Similarly, increased suction width results in earlier separation. The modulation due to increasing suction width is more pronounced than that due to increased APG. Thus, for the levels of APG tested, it can be estimated that the non-uniformity of the APG serves as an indicator of separation onset. That is, regardless of APG magnitude, the narrower the APG width in the spanwise direction, the later the flow will separate.

The behavior of skin friction is more complex near the reattachment point (defined as $C_f = 0 $, $\partial C_f/ \partial x > 0$) than near separation. No immediately discernible trends of reattachment point are readily noticeable from the $C_f$ plot. 
As expected, the magnitude of reverse flow prior to the reattachment point increases as APG increases (for a given suction width). Similarly, the magnitude of forward flow downstream of reattachment increases as APG increases. The same trends are true for increasing suction width at a given APG. Although not as strictly followed as the separation point, the spanwise width of the APG again serves as the better indicator for reverse and forward flow strength at reattachment. For example, the magnitude of reverse flow is greater in all $z_0 = \infty$ cases than all $z_0 = 10\delta$ cases, which in turn are greater than the $z_0 = 5\delta$ cases. The exception is case zinfv065 has a similar magnitude reverse flow to the $z_0 = 10\delta$ cases. 
These insights, in conjunction with the streamline behavior at reattachment shown in Fig. \ref{fig:Uxy}, suggest that three-dimensional LSBs display more gradual reattachment compared to the impinging-type reattachment that occurs in their two-dimensional counterparts \citep{Moin98,Coleman2018,WuMM20}. 

\begin{figure}[h!]
    \centering
    \includegraphics[width=0.49\textwidth]{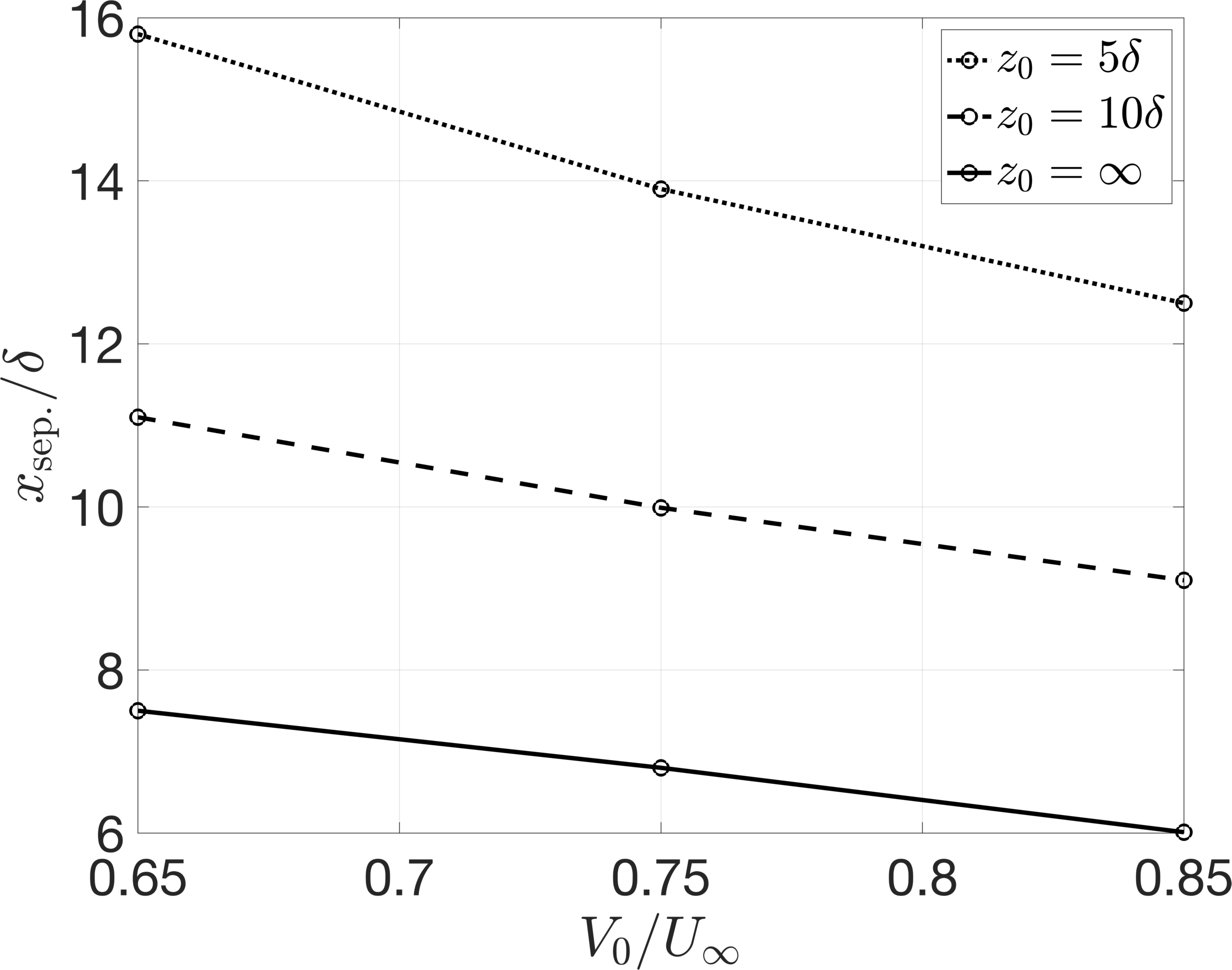}
    \includegraphics[width=0.49\textwidth]{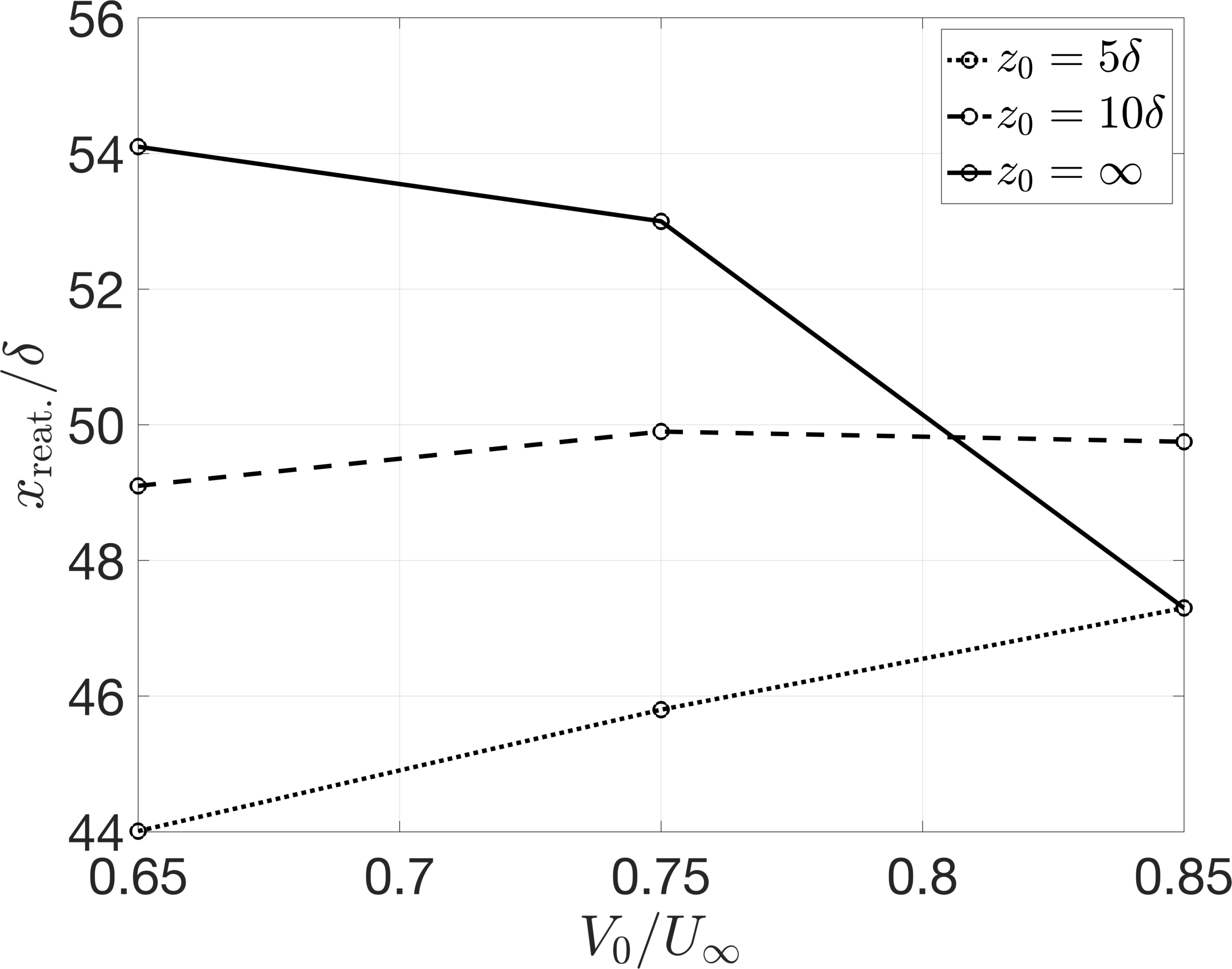}
    \caption{(a) Mean separation location and (b) mean reattachment location as a function of APG strength ($V_0/U_{\infty}$). Note that the flow field is averaged in both spanwise and time. In both plots, line style corresponds to suction width: \dotted $z_0 = 5\delta$; \dashed $z_0 = 10\delta$; \solid $z_0 = \infty$.}
    \label{fig:xsepreat}
\end{figure}

The mean separation and reattachment points (defined above) are plotted as a function of APG strength ($V_0/U_{\infty}$) in Fig. \ref{fig:xsepreat}. 

It is important to note that the separation and reattachment points are calculated using spanwise- and time-averaged streamwise velocity, even though the flow is not separated (or reattached) across the full span in the $z_0\neq \infty$ cases. Thus, Fig. \ref{fig:xsepreat} provides a metric for where the averaged flow in the spanwise direction is separated (or reattached). The behavior of the separation point, as observed in figure \ref{fig:Cf}, is clearly monotonic. That is, the flow separates earlier as APG strength increases and as suction width increases. The behavior of the reattachment point is not as straightforward. With respect to suction strength, two-dimensional LSBs (the solid line) reattach earlier with increasing suction strength. On the contrary, the three-dimensional bubbles reattach later as APG strength increases. With respect to suction width, reattachment occurs earlier as width decreases for the low and moderate APGs. For the largest APG, there is no discernible trend, as the two-dimensional case reattaches earlier than both three-dimensional cases.

\subsection{Turbulence statistics}
\begin{figure}[h!]
    \centering
     \includegraphics[width=0.35\textwidth]{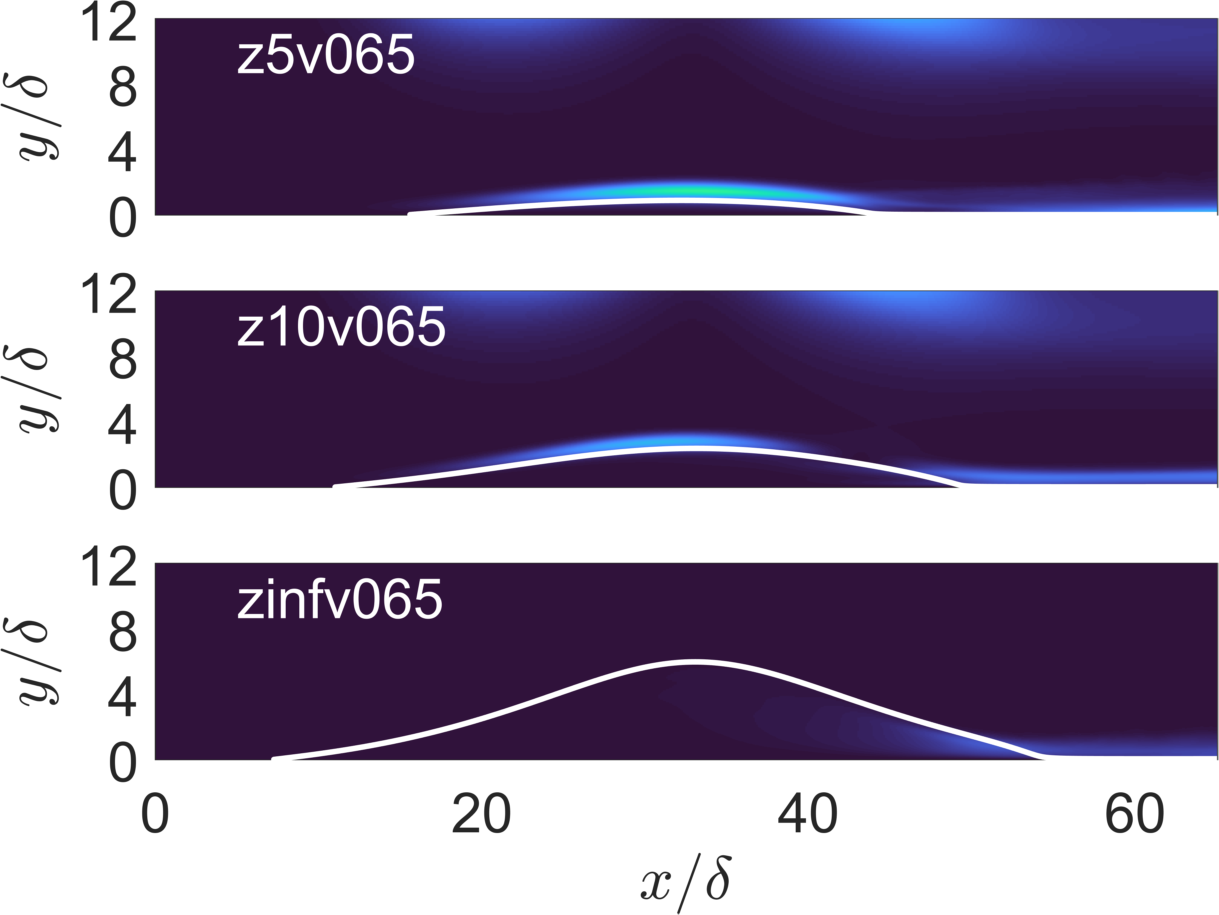}
    \includegraphics[width=0.31\textwidth]{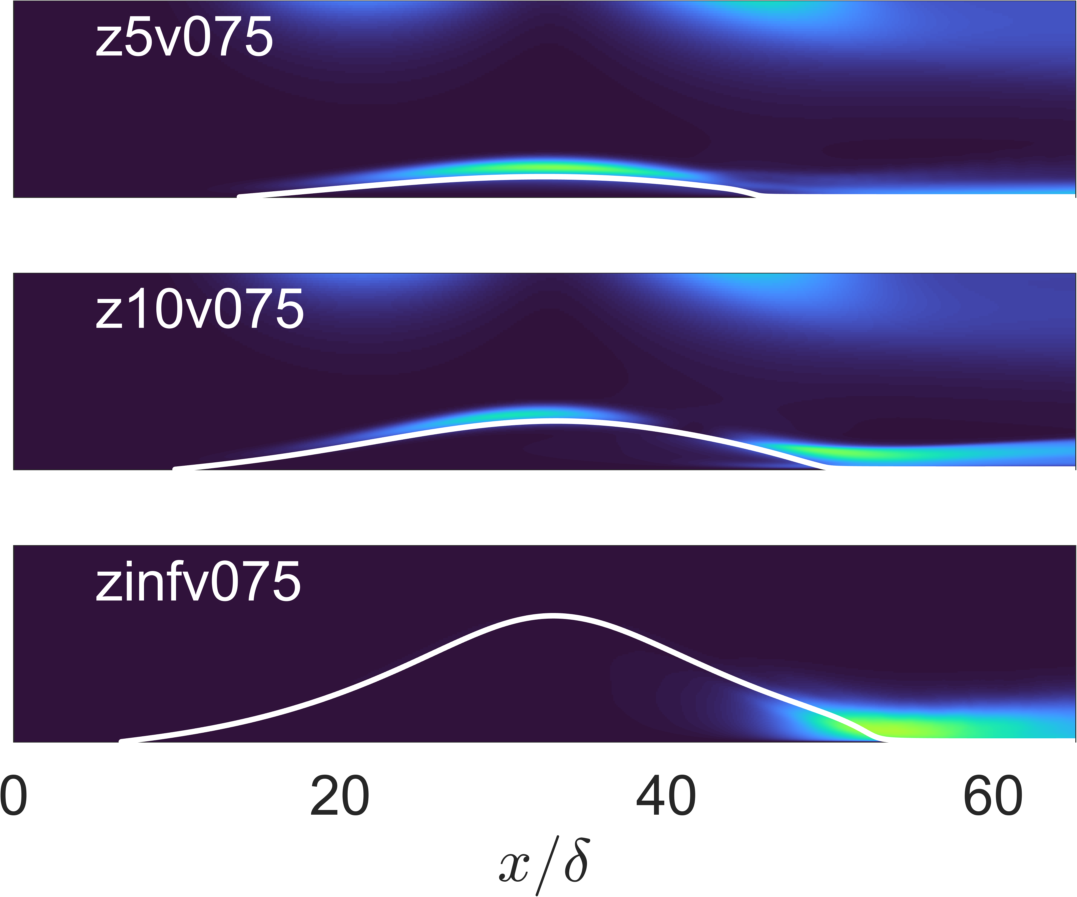}
     \includegraphics[width=0.31\textwidth]{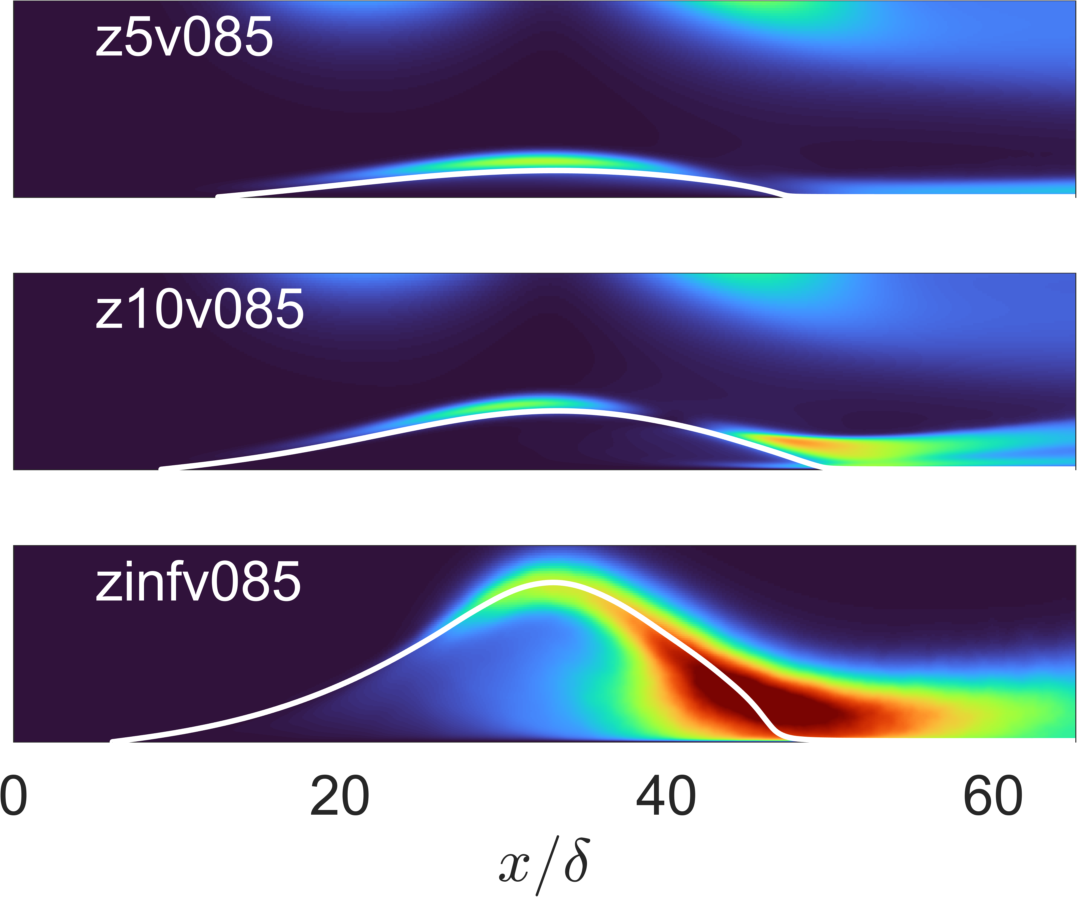}
     \hspace*{0.8cm}
     \includegraphics[width=0.15\textwidth]{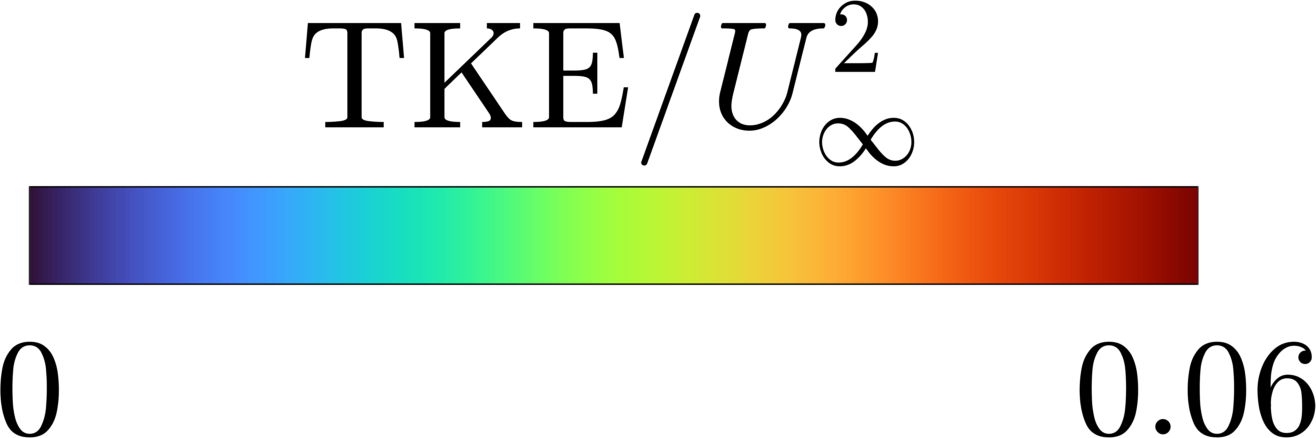}
     \caption{Spanwise averaged turbulence kinetic energy (TKE). The separating streamline is shown by the white solid line.}
     \label{fig:TKE}
\end{figure}

Figure \ref{fig:TKE} shows spanwise-averaged turbulence kinetic energy, defined as $\mathrm{TKE} = 0.5\left < \left(\overline{u'u'} + \overline{v'v'} + \overline{w'w'} \right) \right>$. There are distinct regions of elevated TKE that vary among the cases: 1) the high energy region occurring on the lee side of LSB near the reattachment point, and 2) the elevated region near the crest of the separation region.
Region 1, elevated TKE near reattachment, is a distinguishing feature of canonical separation bubbles, especially LSBs. It is well understood that this region of TKE is a signature of the formation and turbulent breakdown of spanwise roller vortices that form in the shear layer on the rear of the bubble \citep{Marxen12, Yeh19}. As APG or suction width strength increases, the TKE near reattachment increases. 
This indicates that the process of vortex generation, breakdown, and reattachment is influenced by the spanwise extent of the LSB. 
The question remains if spanwise-oriented, roller-type vortices are generated in the three-dimensional cases (and are simply smaller than those of the two-dimensional counterpart) or if the vortex structure of the three-dimensional LSB is fundamentally different from the two-dimensional LSB vortex structure.

The second phenomenon -- enhanced TKE levels near the crest of the LSB -- is not reported in the literature on two-dimensional separation regions. It is present in all non-uniform APG cases. The magnitude of TKE in this region increases as suction width decreases and as suction strength increases. Therefore, case z5v085 experiences the highest levels of TKE on the LSB crest. It should be noted that case zinfv085 is an exception to this trend, however the enhanced TKE on the LSB crest for this case appears to be correlated with the bubble breakdown and reattachment (rather than the standalone phenomena observed in the three-dimensional cases). We will demonstrate this momentarily.
The above observations indicate that there are additional physical mechanisms dictating the evolution and dynamics of three-dimensional LSBs that are not observed in two-dimensional LSBs. To gain further insight into what these mechanisms are, we turn to instantaneous flow field visualizations.

\subsection{Instantaneous flow field}
\begin{figure}[h!]
    \centering
    \includegraphics[width=0.99\textwidth]{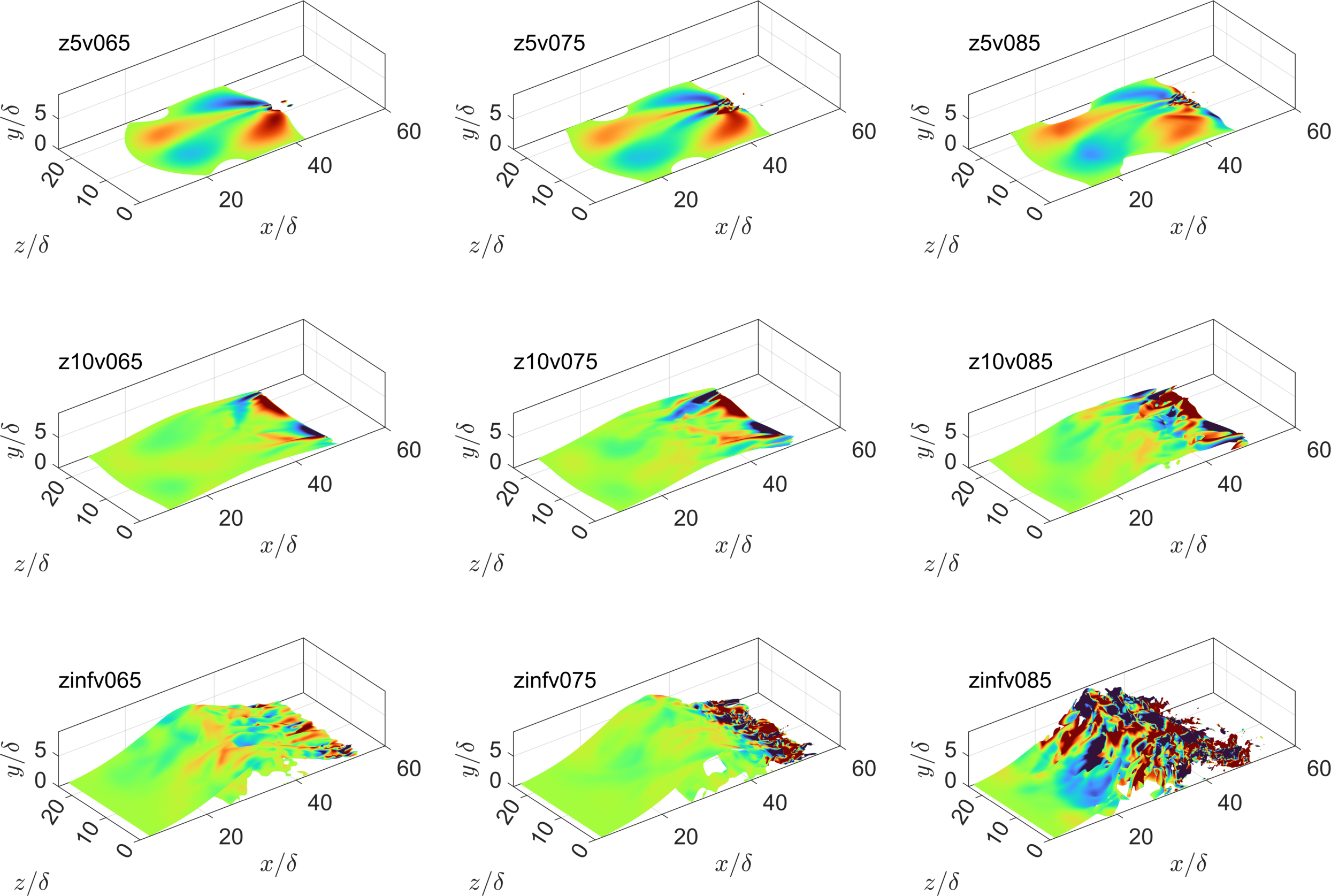}
    \includegraphics[width=0.22\textwidth]{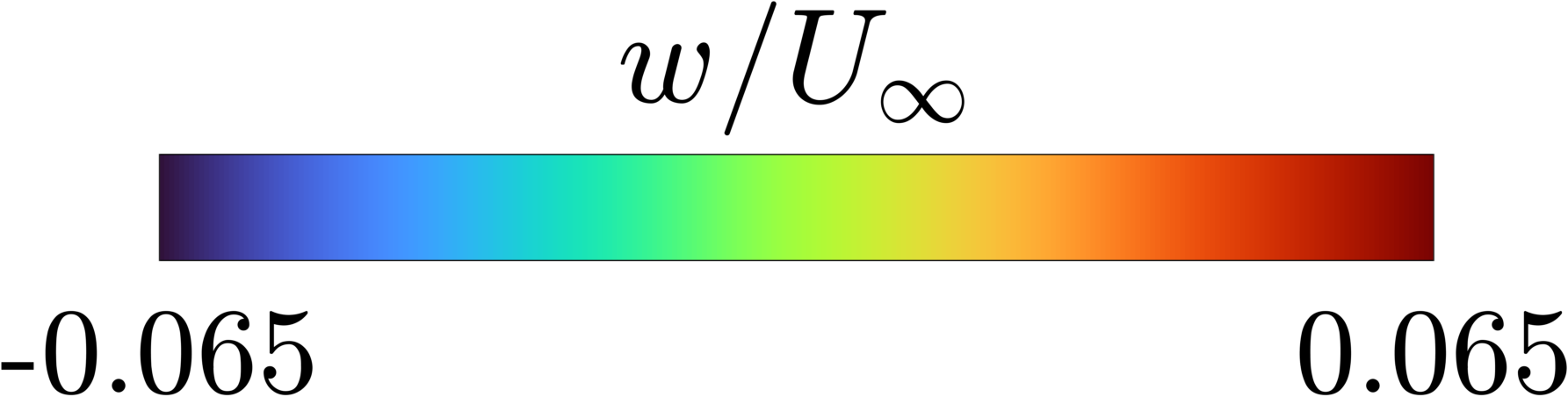}
    \caption{Isosurface of instantaneous streamwise velocity $u=0$. Isosurfaces are colored by instantaneous spanwise velocity ($w/U_{\infty}$). Note that the aspect ratio of the domain is adjusted for clarity and emphasis on the three-dimensional structure.}
    \label{fig:usio3d}
\end{figure}

To provide visualization of the three-dimensional topology of each LSB, the isosurface of instantaneous streamwise velocity $u=0$ is shown in Fig. \ref{fig:usio3d}. The isosurface is colored by instantaneous spanwise velocity ($w/U_{\infty}$). In a qualitative sense, the three-dimensional LSBs are primarily dependent on suction width rather than APG strength. That is, the instantaneous separation region of all three cases at a given suction width appears qualitatively similar in the instantaneous sense. As APG strength increases for a given suction width, the height of the bubble and turbulence intensity at reattachment increases. Therefore, the role of suction width is to dictate the LSB structure, shape, and evolution. Meanwhile, the role of APG strength is determining the height of the bubble, when the transition to turbulence occurs, and the turbulence intensity. For this reason, the instantaneous LSB structure will be discussed with reference to the three suction widths.

We first assess the two-dimensional cases ($z_0 = \infty$). The LSB is initially two-dimensional, as expected.  Downstream, the perturbations grow causing spanwise-variation to the LSB before turbulence ensues. As APG strength increases, this transition occurs earlier and the turbulence level is enhanced (indicated by saturated $w/U_{\infty}$ and fine-scale turbulent structures). The elevated TKE near the crest of the bubble in case zinfv085 (as observed in Fig. \ref{fig:TKE}) is seen here to be a product of a very early transition to intense turbulence.

The upstream portion of LSB for the $z_0=10\delta$ cases appears to be stable and largely two-dimensional. At approximately 75\% of the bubble length, appreciable spanwise fluctuations are present, however the bubble appears to remain primarily laminar, as extremely small-scale structures are not identifiable. The spanwise fluctuations in the rear portion of the LSB appear quasi-symmetric for the low suction strength cases. At reattachment, a strong spanwise current that is quasi-symmetric about the bubble's spanwise centerline forms, indicating fluid moving from the center of the bubble toward the domain boundaries. 

The $z_0=5\delta$ cases display the most anomalous, three-dimensional structure. The leading edge of the LSB is wide, before growing thinner in the mid-section and then wider again in the downstream half. Near separation on the wide leading edge of the LSB, symmetric spanwise flow is observed moving from the bubble centerline toward the outside of the bubble. The wide downstream portion of the LSB displays a highly three-dimensional structure, however is still symmetric about the centerline as the suction boundary at the top. Here, along the crest (i.e., the highest wall-normal portion of the LSB), strong spanwise flow moving away from the bubble centerline is observed.  Nearer to the wall, spanwise flow is seen to bring fluid from the domain boundaries toward the center of the LSB. This spanwise current grows stronger closer to reattachment. Finally at reattachment, a confined region of small-scale turbulent structures exists directly downstream of the LSB crest region. Because this region of small-scale structures is confined directly behind the highest portion of the LSB, this indicates vortices are only shedding from the elevated crest region, and reattachment elsewhere is relatively gradual.

\begin{figure}[h!]
    \centering
    \includegraphics[width=0.8\textwidth]{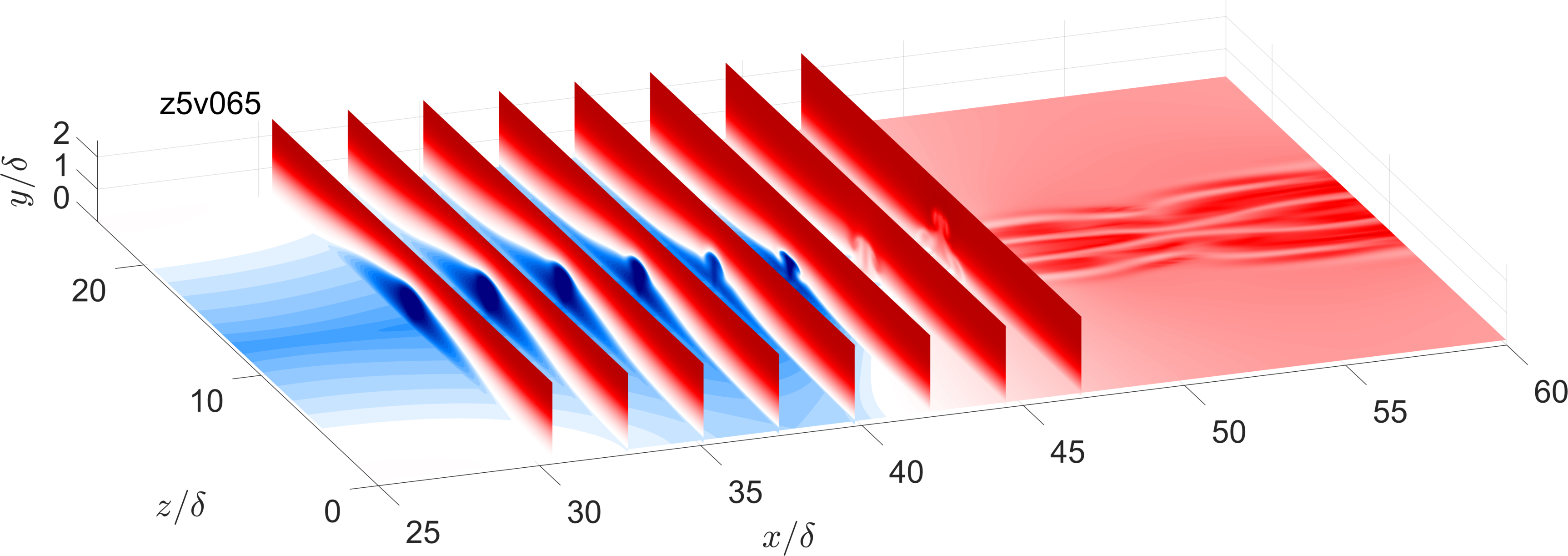}
    \includegraphics[width=0.8\textwidth]{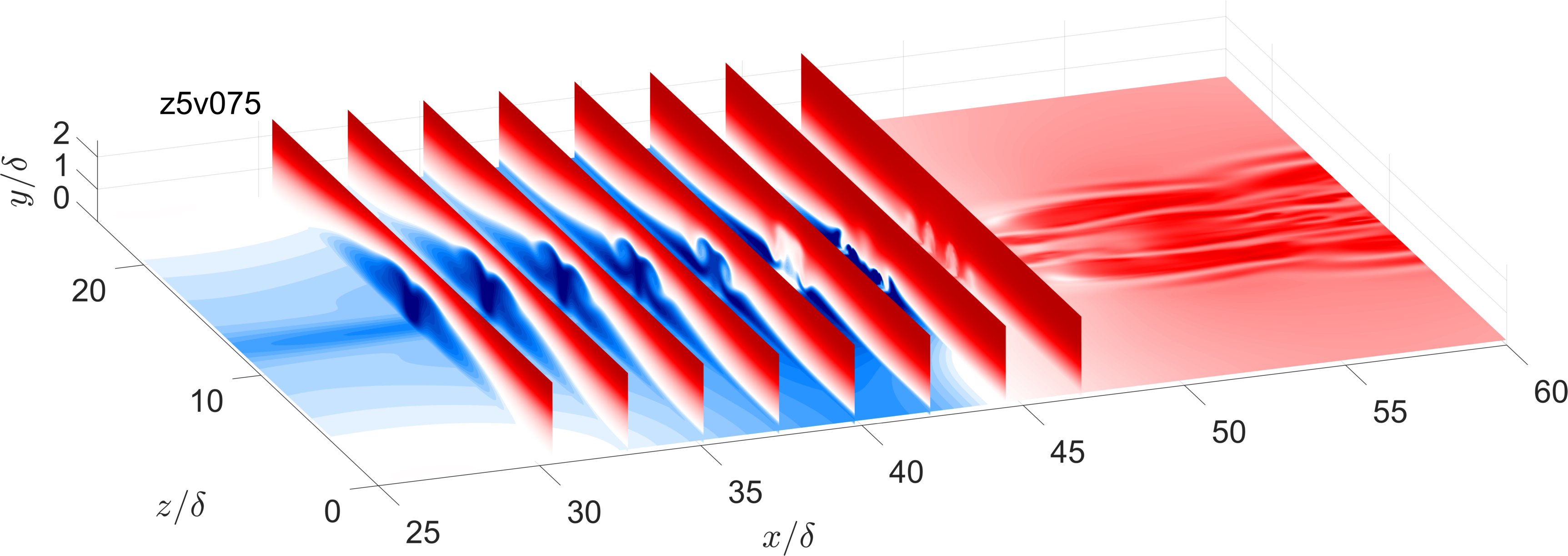}
    \includegraphics[width=0.8\textwidth]{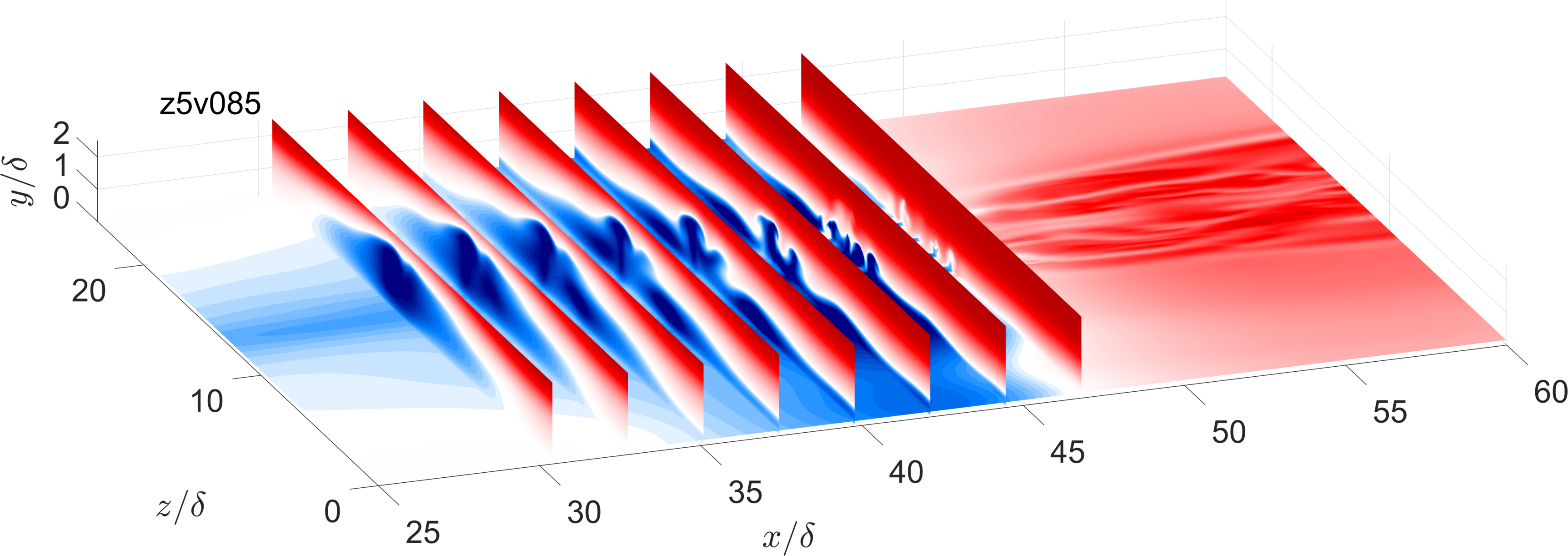}
    \includegraphics[width=0.2\textwidth]{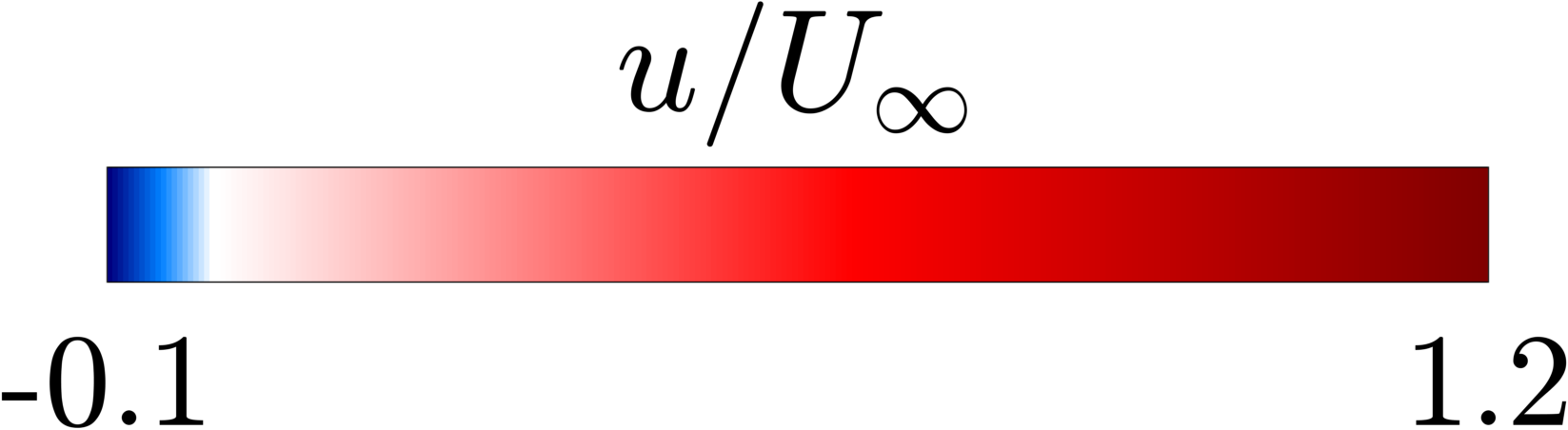}
    \caption{Isometric view of instantaneous streamwise velocity ($u/U_{\infty}$) for the $z_0 = 5\delta$ cases. $z-y$ planes are displayed at $x/\delta = $30.4, 32.7, 35.1, 37.4, 39.7, 42.1, 44.4, and 46.8. The $x-z$ plane is displayed at $y/\delta = 0.14$. From top to bottom: $V_0 = 0.65 U_{\infty}$, $0.75U_{\infty}$, $0.85U_{\infty}$.}
    \label{fig:uinst3d}
\end{figure}

To further investigate the behavior of the downstream portion of the three-dimensional LSBs, the instantaneous streamwise velocity ($u/U_\infty$) downstream of $x/\delta = 25$ is displayed at selected two-dimensional planes in Fig. \ref{fig:uinst3d}. Cross-planes ($z-y$) are visualized from $30.4\delta$ to $46.8\delta$ at a streamwise interval of ~$2.3\delta$. A wall-parallel ($x-z$) plane is visualized at $y/\delta = 0.14$.  Most glaringly, a pair of counter-rotating, streamwise-oriented vortices are observed to form along the crest of the LSB. These vortices are responsible for the enhanced spanwise flow along the crest that was observed in Fig. \ref{fig:usio3d}. This also suggests that the enhanced TKE levels along the crest observed in Fig. \ref{fig:TKE} are due to these structures, as they are most prominent for the $z_0 = 5\delta$ cases. Further downstream, they break down into smaller-scale structures where the flow reattaches. The turbulence induced by this breakdown forms a symmetric wake near the wall that gradually expands in the spanwise direction. The reattachment and wake outside of this region are laminar. This indicates that the small-scale structures observed in Fig. \ref{fig:uinst3d} are due to the spanwise rollers formed on the LSB crest. These rollers are relatively weak in case z5v065. They are larger in case z5v075 and form earlier in the LSB than in case z5v065. Interestingly, a similar structure appears to be present in case z5v085, however it is non-symmetric about the centerline at the depicted moment and is more distorted than in the previous two cases. Similarly, the near-wall turbulent wake is not symmetric about the bubble centerline. It is hypothesized that the strong APG in this case induces greater perturbations to the LSB, disrupting the coherent roller-pair, introducing instantaneous asymmetry to the LSB and wake topology.

\section{Conclusion}

Direct numerical simulations (DNS) are employed to investigate the effects of non-uniform adverse pressure gradients (APGs) on laminar separation bubble (LSB) behavior. Our initial results indicate various qualitative differences between canonical, two-dimensional LSBs and three-dimensional LSBs generated by spanwise inhomogeneous APGs. The suction width is shown to have a more significant impact on both the mean separation point and reattachment  point than suction strength. Additionally, suction width provides a metric for the strength of forward and reverse flow on either side of the reattachment point.  Fundamental qualitative differences are observed in the topology and vortex structure of three-dimensional LSBs in comparison with their two-dimensional counterparts. Rather than the traditional spanwise-oriented roller vortices formed in the shear layer, a streamwise-oriented roller pair along the LSB crest is the dominant structure. The breakdown of this roller-pair appears to drive an isolated region of turbulent reattachment, forming a turbulent wake symmetric about the bubble centerline. The roller-pair leaves a significantly different signature in the turbulence statistics than the spanwise-oriented roller vortices of two-dimensional LSBs. A plethora of open questions arise: what causes this structure to form? Is it a stable, persistent structure or does it break down intermittently? What criteria determine when a distinct roller pair forms? What determines when the roller-pair breaks down at the trailing edge of the LSB? The complete conference paper will address the posed questions, along with a complete analysis of the presented flows.

\section*{Acknowledgments}
The authors gratefully acknowledge the San Diego Supercomputer Center (SDSC) Expanse CPU cluster for computing resources. W.W. acknowledges the support from NSF grant OIA-2131942, monitored by Dr. Hongmei Luo. B.S. acknowledges support from NSF GRFP Award 2235036.  

\bibliography{main}

\end{document}